\documentclass[twocolumn,showpacs,preprintnumbers]{revtex4}
\usepackage{graphicx}
\usepackage{dcolumn}
\usepackage{bm}

\tighten

\begin{document}

\title{Spin and quadrupole correlations in the insulating nematic phase of
spin-1 bosons in optical lattices}
\author{Guang-Ming Zhang$^{1}$ and Lu Yu$^{2}$}
\affiliation{$^{1}$Department of Physics and Center for Advanced Study, Tsinghua
University, Beijing 100084, China;\\
$^{2}$Institute of Theoretical Physics and Interdisciplinary Center of
Theoretical Studies, CAS, Beijing 100080, China.}
\date{\today}

\begin{abstract}
We consider the effective model of $H=-J_{1}\sum_{<i,j>}\mathbf{S}_{i}\cdot
\mathbf{S}_{j}-J_{2}\sum_{<i,j>}\left( \mathbf{S}_{i}\cdot \mathbf{S}%
_{j}\right) ^{2}$, describing the Mott insulating phase with odd number of
spin-1 bosons in optical lattices ($J_{2}>J_{1}>0$). In terms of an SU(3)
boson representation, a valence bond mean field theory is developed. In 1D,
a \textit{first-order} quantum phase transition from a spin singlet to a
spin nematic phase with \textit{gapful} excitations is identified at $%
J_{1}/J_{2}=0.19833$ , while on a 2D square lattice a spin nematic ordered
phase with gapless excitations prevails. In both 1D and 2D cases, we predict
that the spin structure factor displays dominant antiferromagnetic
fluctuations, while the quadrupole structure factor exhibits strong
ferroquadrupolar correlations.
\end{abstract}

\maketitle

Degenerate alkali atoms are considered weakly interacting boson gas due to
the smallness of the scattering length compared with the inter-particle
separation. However, the situation changes dramatically when an optical
potential created by standing laser beams confines particles in valleys of
the periodic potential and strongly enhances local interactions. Recently,
the nontrivial Mott insulating state of bosonic atoms in optical lattices
has been demonstrated experimentally \cite{greiner}. In addition to solid
state systems, spinor atoms in optical lattices provide a novel realization
of quantum magnetic systems with the possibility to tune\ various parameters
of the effective models in the absence of disorder \cite{jaksch,duan,cirac}.

Alkali atoms have a nuclear spin 3/2. Lower energy hyperfine manifold has
three magnetic sublevels and a total moment $S=1$. In order to observe the
quantum spin phenomena experimentally, one has to consider a system with
\textit{small} number of particles and \textit{strong} interactions. In the
insulating state, atoms are localized, and fluctuations in the particle
number on each site are suppressed. Virtual tunnelling of atoms between
neighboring sites induces effective spin interactions, leading to novel
quantum magnetic phases \cite{demler-zhou,zhou,demler,skyip}.

The boson-Hubbard model is used to describe the low-energy physics of spin-1
bosonic atoms in an optical lattice \cite{ho},%
\begin{eqnarray}
H &=&-t\sum_{<i,j>,m}(a_{i,m}^{\dagger }a_{j,m}+h.c.)+\frac{U_{0}}{2}%
\sum_{i}n_{i}(n_{i}-1)  \nonumber \\
&&+\frac{U_{2}}{2}\sum_{i}(\mathbf{S}_{i}^{2}-2n_{i})-\mu \sum_{i}n_{i},
\end{eqnarray}%
where $n_{i}=\sum_{m}a_{i,m}^{\dagger }a_{i,m}$ and $S_{i}^{\alpha
}=\sum_{m,n}a_{i,m}^{\dagger }T_{m,n}^{\alpha }a_{i,n}$ denote the number of
atoms and the spin operator on the $i$ site ($m=-1,0,1$), respectively,
while $T^{\alpha }$ denote the matrices for the spin-1 particles.

In the $t=0$ limit with \textit{odd} number of bosons per site, the bosonic
symmetry of the wave function requires that each site has a localized spin
with \textit{odd} quantum numbers. The interactions of Eq.(1) are minimized
when the localized spins take the smallest possible value $S=1$. Away from
this limit, the second order perturbation theory in $t$ leads to an
effective spin superexchange model \cite{zhou,demler,tsuchiya}%
\begin{equation}
H=-J_{1}\sum_{<i,j>}\mathbf{S}_{i}\cdot \mathbf{S}_{j}-J_{2}\sum_{<i,j>}%
\left( \mathbf{S}_{i}\cdot \mathbf{S}_{j}\right) ^{2},
\end{equation}%
where the coupling parameters are confined to $0<J_{1}/J_{2}<1$. For a
two-spin problem, the first term favors a polarized configuration, while the
second prefers a spin singlet form. Thus, the effective model acts as a
spin-1 Heisenberg \textit{antiferromagnet} with ferromagnetic frustrations.

It has been an outstanding issue to study the ground states of this
effective model \cite{chen-levy,harada,zhou,demler,skyip}. More exotic
ground states may be realized in 1D \cite{chubukov,solyom,skyip}, but there
is no consensus yet. In this paper, using an SU(3) boson representation, we
develop a valence bond mean field (MF) theory. In 1D, a \textit{first-order}
quantum phase transition is found from a spin singlet ($J_{1}/J_{2}<0.19833$%
) to a (short-range ordered) spin nematic phase ($J_{1}/J_{2}>0.19833$) with
gapful excitations, while on a 2D square lattice an \textit{ordered} spin
nematic phase is always obtained. Moreover, the spin and quadrupole
correlation spectra in both disordered and ordered nematic phases are
calculated explicitly, to be checked by future experiments.

To describe the spin-1 operators, an SU(3) bosonic representation is defined
by generators $F_{m}^{n}(i)=a_{i,m}^{\dagger }a_{i,n}$, where indices $m$
and $n$ specify the spin projection with values $-1,0,1$. The commutation
relation is satisfied
\begin{equation}
\left[ F_{m}^{n}(i),F_{\mu }^{\nu }(j)\right] =\delta _{i,j}\left[ \delta
_{n,\mu }F_{m}^{\nu }(i)-\delta _{m,\nu }F_{\mu }^{n}(i)\right] ,
\end{equation}%
which forms an SU(3) Lie algebra. The corresponding spin operators are
expressed as $S_{i}^{+}=\sqrt{2}(a_{i,0}^{\dagger }a_{i,-1}+a_{i,1}^{\dagger
}a_{i,0})$, $S_{i}^{-}=(S_{i}^{+})^{\dagger }$, $S_{i}^{z}=(a_{i,1}^{\dagger
}a_{i,1}-a_{i,-1}^{\dagger }a_{i,-1})$. With this boson representation, we
can verify $\left[ S_{i}^{+},S_{j}^{-}\right] =2S_{i}^{z}\delta _{i,j},$ $%
\left[ S_{i}^{z},S_{j}^{\pm }\right] =\pm S_{i}^{\pm }\delta _{i,j}$. In
order to fix $S_{i}^{2}=S(S+1)=2$, a local constraint $\sum_{m}a_{i,m}^{%
\dagger }a_{i,m}=1$ has to be imposed.

To make the spin representation symmetric, linear combinations are
introduced: $b_{i,1}=\frac{1}{\sqrt{2}}\left( a_{i,-1}-a_{i,1}\right) $, $%
b_{i,2}=\frac{-i}{\sqrt{2}}\left( a_{i,-1}+a_{i,1}\right) $, and $%
b_{i,3}=a_{i,0}$, and the spin (dipolar) operators are written in
antisymmetric form
\begin{eqnarray}
S_{i}^{x} &=&-i(b_{i,2}^{\dagger }b_{i,3}-b_{i,3}^{\dagger }b_{i,2}),
\nonumber \\
S_{i}^{y} &=&-i(b_{i,3}^{\dagger }b_{i,1}-b_{i,1}^{\dagger }b_{i,3}),
\nonumber \\
S_{i}^{z} &=&-i(b_{i,1}^{\dagger }b_{i,2}-b_{i,2}^{\dagger }b_{i,1}),
\end{eqnarray}%
and the local constraint holds as $\sum_{\alpha }b_{i,\alpha }^{\dagger
}b_{i,\alpha }=1$. Actually for spin-1 bosonic atoms, \textit{quadrupole}
operators can be defined by%
\begin{eqnarray}
Q_{i}^{(0)} &=&3(S_{i}^{z})^{2}-2=b_{i,1}^{\dagger }b_{i,1}+b_{i,2}^{\dagger
}b_{i,2}-2b_{i,3}^{\dagger }b_{i,3},  \nonumber \\
Q_{i}^{(2)} &=&(S_{i}^{x})^{2}-(S_{i}^{y})^{2}=-(b_{i,1}^{\dagger
}b_{i,1}-b_{i,2}^{\dagger }b_{i,2}),  \nonumber \\
Q_{i}^{xy} &=&S_{i}^{x}S_{i}^{y}+S_{i}^{y}S_{i}^{x}=-(b_{i,1}^{\dagger
}b_{i,2}+b_{i,2}^{\dagger }b_{i,1}),  \nonumber \\
Q_{i}^{xz} &=&S_{i}^{x}S_{i}^{z}+S_{i}^{z}S_{i}^{x}=-(b_{i,1}^{\dagger
}b_{i,3}+b_{i,3}^{\dagger }b_{i,1}),  \nonumber \\
Q_{i}^{yz} &=&S_{i}^{y}S_{i}^{z}+S_{i}^{z}S_{i}^{y}=-(b_{i,2}^{\dagger
}b_{i,3}+b_{i,3}^{\dagger }b_{i,2}).
\end{eqnarray}%
Three dipole and five quadrupole operators form generators of the SU(3) Lie
group, as in the Gell-Mann matrix representation. Then, the effective spin
model is expressed as:%
\begin{eqnarray}
H &=&-\sum_{<i,j>}\sum_{\alpha ,\beta }\left[ J_{1}b_{i,\alpha }^{\dagger
}b_{i,\beta }b_{j,\beta }^{\dagger }b_{j,\alpha }\right.  \nonumber \\
&&\left. +(J_{2}-J_{1})b_{i,\alpha }^{\dagger }b_{i,\beta }b_{j,\alpha
}^{\dagger }b_{j,\beta }\right] .
\end{eqnarray}%
In the limit $J_{1}=J_{2}$, the model is reduced to an SU(3) symmetric
\textit{ferromagnetic} superexchange model, invariant under the uniform
SU(3) transformation, for which the spin ferromagnetic and spin nematic
long-range order can coexist in higher dimensions \cite{batista}. On the
other hand, in the limit $J_{1}=0$, the model is reduced to an SU(3)
symmetric valence-bond \textit{antiferromagnetic} model, invariant under the
staggered conjugate transformations of the two sublattices, leading to a
spin dimerization in 1D \cite{affleck,barber}.

It is known that an SU(2) Schwinger boson MF theory can describe rather well
the spin correlations for spin-1/2 Heisenberg antiferromagnets \cite%
{arovas-auerbach}. To develop a similar valence bond MF theory, we introduce
symmetric pairing parameters%
\begin{eqnarray}
\Delta _{\alpha ,\alpha } &=&-\langle b_{i,\alpha }^{\dagger }b_{j,\alpha
}^{\dagger }\rangle ,  \nonumber \\
\Delta _{1,2} &=&-\left\langle b_{i,1}^{\dagger }b_{j,2}^{\dagger
}+b_{i,2}^{\dagger }b_{j,1}^{\dagger }\right\rangle ,  \nonumber \\
\Delta _{1,3} &=&-\left\langle b_{i,1}^{\dagger }b_{j,3}^{\dagger
}+b_{i,3}^{\dagger }b_{j,1}^{\dagger }\right\rangle ,  \nonumber \\
\Delta _{2,3} &=&-\left\langle b_{i,2}^{\dagger }b_{j,3}^{\dagger
}+b_{i,3}^{\dagger }b_{j,2}^{\dagger }\right\rangle .
\end{eqnarray}%
To preserve the SU(2) spin rotational symmetry of the model, we have to
assume $\Delta _{1,1}=\Delta _{2,2}=\Delta _{3,3}\equiv \Delta _{a}$ and $%
\Delta _{1,2}=\Delta _{1,3}=\Delta _{2,3}\equiv \Delta _{b}$. By introducing
a Nambu spinor $\Psi _{\mathbf{k}}=(b_{\mathbf{k},1}^{\dagger },b_{\mathbf{k}%
,2}^{\dagger },b_{\mathbf{k},3}^{\dagger },b_{-\mathbf{k},1},b_{-\mathbf{k}%
,2},b_{-\mathbf{k},3})$, we can rewrite the MF Hamiltonian in a matrix form
\begin{eqnarray}
H_{mf} &=&\frac{1}{2}\sum_{\mathbf{k}}\Psi _{\mathbf{k}}^{\dagger }\mathbf{H}%
_{mf}(\mathbf{k})\Psi _{\mathbf{k}}  \nonumber \\
&&+\frac{3}{2}\left[ \left( 3J_{2}-J_{1}\right) \Delta _{a}^{2}+J_{1}\Delta
_{b}^{2}\right] Nz-\frac{5}{2}\lambda N,
\end{eqnarray}%
where the local constraint has been implemented by a Lagrangian multiplier $%
\lambda $, $N$ the total number of lattice sites, and $z$ the number of the
nearest neighbor sites. The MF Hamiltonian matrix is given by
\begin{equation}
\mathbf{H}_{mf}(\mathbf{k})=\lambda +2\Delta _{a}(\mathbf{k})\sigma
_{x}\otimes I+\Delta _{b}(\mathbf{k})\sigma _{x}\otimes M,
\end{equation}%
where $M=\left(
\begin{array}{ccc}
0 & 1 & 1 \\
1 & 0 & 1 \\
1 & 1 & 0%
\end{array}%
\right) $, $I$ is the $3\times 3$ unit matrix, $\sigma _{\alpha }$ $(\alpha
=x,y,z)$ are Pauli matrices, $\Delta _{a}(\mathbf{k})=\frac{z}{2}\Delta
_{a}(3J_{2}-J_{1})\gamma _{_{\mathbf{k}}}$, $\Delta _{b}(\mathbf{k})=z\Delta
_{b}J_{1}\gamma _{_{\mathbf{k}}}$, $\gamma _{_{\mathbf{k}}}=\frac{1}{d}%
\sum_{\delta }\cos \mathbf{k\cdot \delta }$, and $\mathbf{\delta }$ the
nearest neighbor vector. The corresponding Matsubara Green function (GF) is
thus deduced to
\begin{equation}
\mathbf{G}^{-1}(\mathbf{k},i\omega _{n})=i\omega _{n}\sigma _{z}\otimes I-%
\mathbf{H}_{mf}(\mathbf{k}),
\end{equation}%
where $\omega _{n}$ is the bosonic Matsubara frequency. The poles of the GF
matrix give rise to the quasiparticle spectra: $\epsilon _{1}(\mathbf{k})=%
\sqrt{\lambda ^{2}-4\left[ \Delta _{a}(\mathbf{k})+\Delta _{b}(\mathbf{k})%
\right] ^{2}}$ and $\epsilon _{2}(\mathbf{k})=\sqrt{\lambda ^{2}-\left[
2\Delta _{a}(\mathbf{k})-\Delta _{b}(\mathbf{k})\right] ^{2}}$, where the
lower band $\epsilon _{1}(\mathbf{k})$ is singly occupied while the higher
band $\epsilon _{2}(\mathbf{k})$ is doubly degenerate.

From the free energy of the system, the saddle point equations at $T=0$K are
derived as%
\begin{eqnarray}
&&\frac{1}{N}\sum_{\mathbf{k}}\left[ \frac{\lambda }{2\epsilon _{1}(\mathbf{k%
})}+\frac{\lambda }{\epsilon _{2}(\mathbf{k})}\right] =\frac{5}{2},
\nonumber \\
&&\frac{1}{N}\sum_{\mathbf{k}}\frac{\gamma _{\mathbf{k}}^{2}}{\epsilon _{1}(%
\mathbf{k})}=\frac{2\left( \Delta _{a}+\Delta _{b}\right) }{z\left[
(3J_{2}-J_{1})\Delta _{a}+2J_{1}\Delta _{b}\right] },  \nonumber \\
&&\frac{1}{N}\sum_{\mathbf{k}}\frac{\gamma _{\mathbf{k}}^{2}}{\epsilon _{2}(%
\mathbf{k})}=\frac{2\Delta _{a}-\Delta _{b}}{z\left[ (3J_{2}-J_{1})\Delta
_{a}-J_{1}\Delta _{b}\right] },
\end{eqnarray}%
while the ground state energy is $E_{g}=-\frac{3}{2}zN\left[
(3J_{2}-J_{1})\Delta _{a}^{2}+J_{1}\Delta _{b}^{2}\right] $. From the GF
matrix, we can also derive the double-time GFs of the boson operators%
\begin{eqnarray}
&&\langle \langle b_{\mathbf{k},2}|b_{\mathbf{k},1}^{\dagger }\rangle
\rangle =\langle \langle b_{\mathbf{k},3}|b_{\mathbf{k},1}^{\dagger }\rangle
\rangle =\langle \langle b_{\mathbf{k},3}|b_{\mathbf{k},2}^{\dagger }\rangle
\rangle  \nonumber \\
&=&\frac{-\left( i\omega _{n}+\lambda \right) \left[ 4\Delta _{a}(\mathbf{k}%
)\Delta _{b}(\mathbf{k})+\Delta _{b}^{2}(\mathbf{k})\right] }{\left[ \omega
_{n}^{2}+\epsilon _{1}^{2}(\mathbf{k})\right] \left[ \omega
_{n}^{2}+\epsilon _{2}^{2}(\mathbf{k})\right] },
\end{eqnarray}%
and $\langle \langle b_{\mathbf{k},1}|b_{\mathbf{k},1}^{\dagger }\rangle
\rangle =\langle \langle b_{\mathbf{k},2}|b_{\mathbf{k},2}^{\dagger }\rangle
\rangle =\langle \langle b_{\mathbf{k},3}|b_{\mathbf{k},3}^{\dagger }\rangle
\rangle $. Using the spectral representation, we can calculate the
expectation values at $T=0$K,%
\begin{eqnarray}
\langle b_{i,1}^{\dagger }b_{i,2}\rangle &=&\langle b_{i,1}^{\dagger
}b_{i,3}\rangle =\langle b_{i,2}^{\dagger }b_{i,3}\rangle  \nonumber \\
&=&\frac{1}{6N}\sum_{\mathbf{k}}\left[ \frac{\lambda }{\epsilon _{1}(\mathbf{%
k})}-\frac{\lambda }{\epsilon _{2}(\mathbf{k})}\right] ,
\end{eqnarray}%
and $\langle b_{i,1}^{\dagger }b_{i,1}\rangle =\langle b_{i,2}^{\dagger
}b_{i,2}\rangle =\langle b_{i,3}^{\dagger }b_{i,3}\rangle =1/3$. Therefore,
we find $\langle S_{i}^{\alpha }\rangle =0$ and $\langle \left(
S_{i}^{\alpha }\right) ^{2}\rangle =2/3$ ($\alpha =x,y,z$), exhibiting that
both time reversal and spin rotational symmetries are well preserved in the
present MF theory. Then the spin nematic state with a quadrupole moment $%
\langle Q_{i}^{xy}\rangle =\langle Q_{i}^{xz}\rangle =\langle
Q_{i}^{yz}\rangle \equiv -Q$ is a possible ground state.

To evaluate the spin spatial correlations in the Mott insulating phase, the
dynamic correlation functions should be calculated. By expressing the spin
operators in terms of the Nambu spinor, the spin correlations are given by
\begin{eqnarray}
&&\chi ^{\alpha ,\beta }(\mathbf{q},i\omega _{m})  \nonumber \\
&=&\frac{1}{4\beta N}\sum_{\mathbf{k},\omega _{n}}\mathrm{Tr}\left[ \Gamma
_{\alpha }\mathbf{G}(\mathbf{k},i\omega _{n})\Gamma _{\beta }\mathbf{G}(%
\mathbf{k+q},i\omega _{n}+i\omega _{m})\right] ,  \nonumber
\end{eqnarray}%
where $\Gamma _{x}$, $\Gamma _{y}$, $\Gamma _{z}$ denote the corresponding $%
6\times 6$ matrices of $S_{i}^{x}$, $S_{i}^{y}$, $S_{i}^{z}$, respectively,
and both $\omega _{n}$ and $\omega _{m}$ are bosonic Matsubara frequencies.
Inserting the GF matrix and tracing over matrices, we find%
\begin{equation}
\chi ^{x,x}(\mathbf{q},i\omega _{m})=\chi ^{y,y}(\mathbf{q},i\omega
_{m})=\chi ^{z,z}(\mathbf{q},i\omega _{m}),
\end{equation}%
displaying an SU(2) spin rotational symmetry. The corresponding imaginary
part $\chi ^{\prime \prime }(\mathbf{q},\omega )$ can also be obtained
through performing the summation over the Matsubara frequency and analytic
continuation \cite{gmzhang}. According to the fluctuation-dissipation
theorem, the static spin structure factor is thus obtained $S_{D}\left(
\mathbf{q}\right) =\frac{-1}{\pi }\int_{-\infty }^{+\infty }d\omega \left[
1+n_{B}(\omega )\right] \chi ^{\prime \prime }(\mathbf{q},\omega )$.
Furthermore, when the correlation function of the quadrupole operator is
defined by $S_{Q}(\mathbf{r-r}^{\prime })=\langle \lbrack 3\left( S_{\mathbf{%
r}}^{z}\right) ^{2}-2][3\left( S_{\mathbf{r}^{\prime }}^{z}\right)
^{2}-2]\rangle $, the static quadrupole structure factor $S_{Q}\left(
\mathbf{q}\right) $ can be evaluated as well \cite{gmzhang}.

In 1D, $z=2$ and $\gamma _{\mathbf{k}}=\cos k$. For a given value of $%
J_{1}/J_{2}$, the saddle point equations are solved numerically, results
being displayed in Fig.1. For $0\leq J_{1}/J_{2}<0.19833$, we find $\Delta
_{b}=0$, $\Delta _{a}=0.5077J_{2}$, the quasiparticle band $\epsilon _{1}(k)$
and $\epsilon _{2}(k)$ are degenerate,\ so the ground state is a spin
singlet with an energy gap. From the point of view of spin correlations,
such a spin singlet phase is similar to the spin dimerized state in the
limit $J_{1}=0$ \cite{barber}. For $J_{1}/J_{2}>0.19833$, both $\Delta _{b}$
and $\Delta _{a}$ are finite, and two energy gaps in the quasiparticle bands
are found $\Delta _{s,1}=\sqrt{\lambda ^{2}-4\left[ (3J_{2}-J_{1})\Delta
_{a}+2J_{1}\Delta _{b}\right] ^{2}}$ and $\Delta _{s,2}=\sqrt{\lambda ^{2}-4%
\left[ (3J_{2}-J_{1})\Delta _{a}-J_{1}\Delta _{b}\right] ^{2}}$ at momenta $%
k=0$, $\pm \pi $. Moreover, the quadrupole moment $Q$ jumps from $0$ to $%
0.16897$ at the critical coupling $J_{1}/J_{2}=0.19833$, and then the system
is in a short-range ordered spin singlet phase with a \textit{finite}
quadrupole moment. Thus, there is a \textit{first-order} quantum phase
transition from the spin singlet to spin nematic phase with gapful
excitations.

\begin{figure}[tbp]
\begin{center}
\includegraphics[width=2.3in]{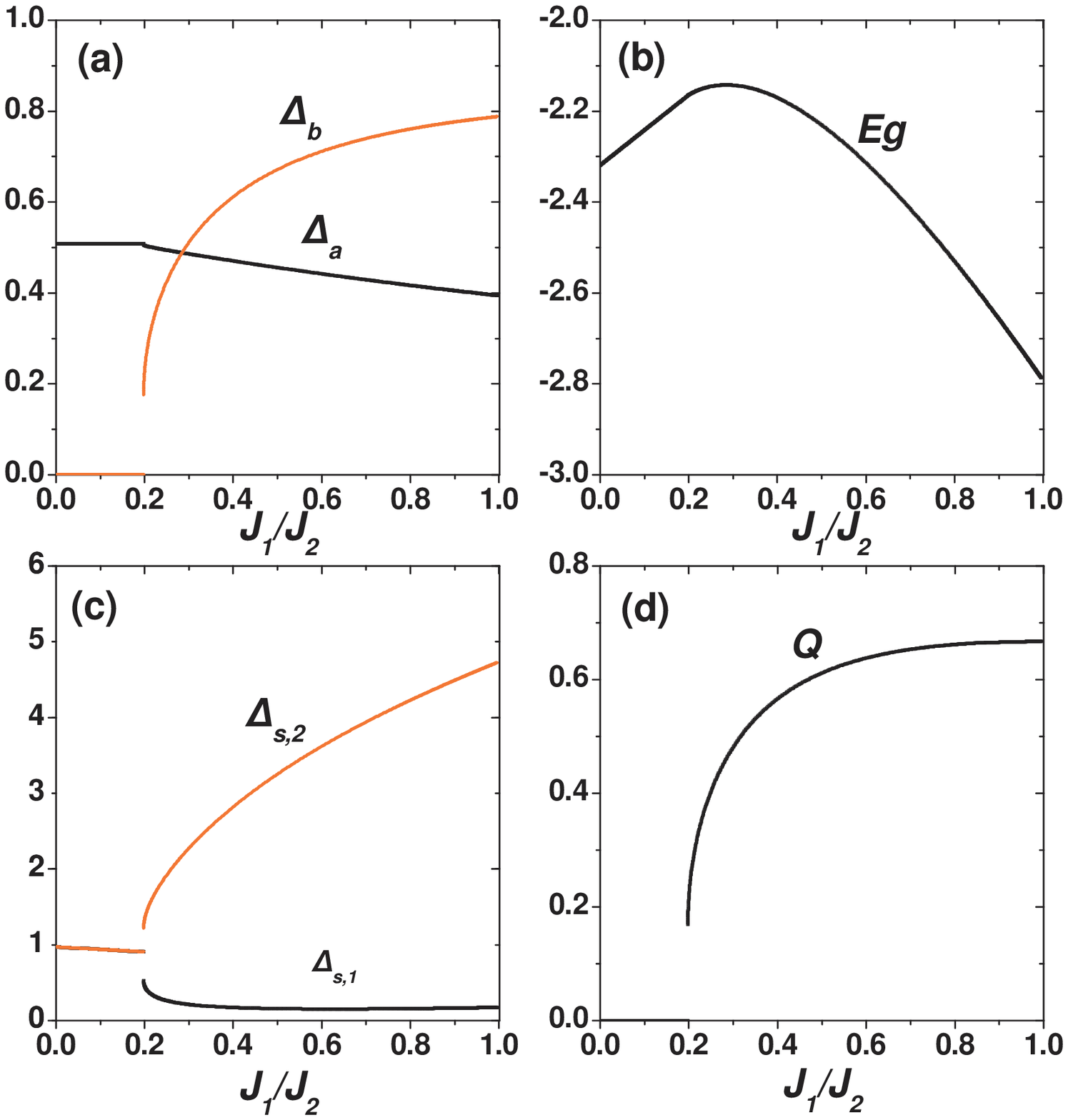}
\end{center}
\caption{Saddle point solution in 1D is displayed: the valence bond
parameters (a), the ground state energy (b), two energy gaps of
quasiparticle bands (c), and the quadrupole moment (d) as functions of $%
J_1/J_2$. $J_{2}$ is taken as the energy unit.}
\label{fig1}
\end{figure}

If we rewrite $J_{1}=-\cos \theta $ and $J_{2}=-\sin \theta $, the critical
coupling corresponds to $\theta _{c}=-0.5633\pi $. In fact, a \textit{%
disordered} spin nematic (non-dimerized) phase \textit{with breaking the
SU(2) spin rotational symmetry} was first suggested around $\theta \approx
-3\pi /4$ by Chubukov \cite{chubukov}. However, subsequent numerical work %
\cite{solyom} did not support this proposal and it was then believed that
the dimerized phase prevails in the whole regime $-3\pi /4\leq \theta \leq
-\pi /2$, i.e., $0<J_{1}/J_{2}<1$. On the other hand, recent quantum Monte
Carlo \cite{kawashima}, density matrix renormalization group calculations %
\cite{lauchli}, and quantum field theory approach \cite{ivanov} indicate
that the dimerized phase may end at $\theta \sim -0.67\pi $, casting doubts
on the earlier anticipations.

To put our MF results on a solid ground, the static spin structure factor $%
S_{D}(q)$ and quadrupole structure factor $S_{Q}(q)$ are calculated at $T=0$%
K and displayed in Fig.2. We find that $S_{D}(q)$ exhibits a broad peak
around the antiferromagnetic wave vector $q=\pi $ in both spin singlet and
gapped spin nematic phase. However, the quadrupole structure factor $%
S_{Q}(q) $ shows a broad peak around the ferromagnetic wave vector $q=0$ in
the spin singlet phase, while $S_{Q}(q)$ exhibits a \textit{sharp} resonance
at $q=0$ in the spin nematic phase, indicating strong \textit{%
ferroquadrupolar} spatial correlations.

\begin{figure}[tbp]
\begin{center}
\includegraphics[width=2.1in]{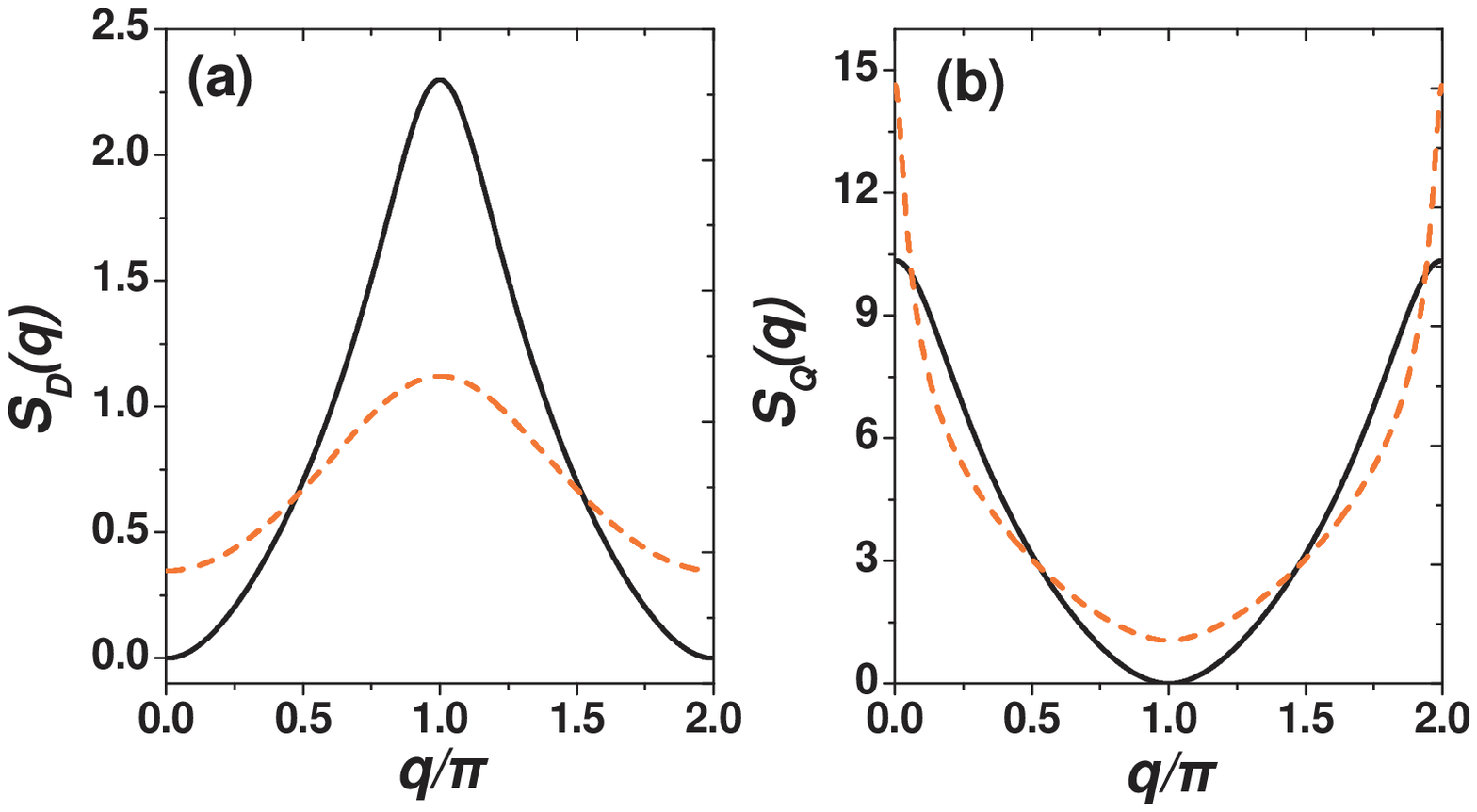}
\end{center}
\caption{Spin structure factor (a) and quadrupole structure factors (b) in
1D for spin singlet phase (solid line $J_{1}/J_{2}=0.1$) and spin nematic
phase (dashed line $J_{1}/J_{2}=0.5$). }
\label{fig2}
\end{figure}

On a 2D square lattice, we have $z=4$ and $\gamma _{\mathbf{k}}=(\cos
k_{x}+\cos k_{y})/2$. At $T=0$K, the conversion\ from summation over
momentum to integral will be invalid as $\lambda \rightarrow 4\left[%
(3J_{2}-J_{1})\Delta _{a}+2J_{1}\Delta _{b}\right] $, and then the
quasiparticle band $\epsilon _{1}(\mathbf{k})$ becomes gapless and linear
near $\mathbf{k}^{\ast }=(0,0)$ and $(\pi ,\pi )$. The Bose-Einstein
condensation thus occurs, while a finite energy gap still exists in the band
$\epsilon _{2}(\mathbf{k})$. By separating the divergent term from the
summation, we can introduce a superfluid density $\rho _{s}$ and then solve
the saddle point equations. The results are shown in Fig.3, where a finite
quadruple moment $Q$ is always obtained and it increases as the coupling $%
J_{1}/J_{2}$ grows. The ground state in the the regime $0\leq
J_{1}/J_{2}\leq 1$ is thus a long-range \textit{ordered} spin nematic phase.

\begin{figure}[tbp]
\begin{center}
\includegraphics[width=2.1in]{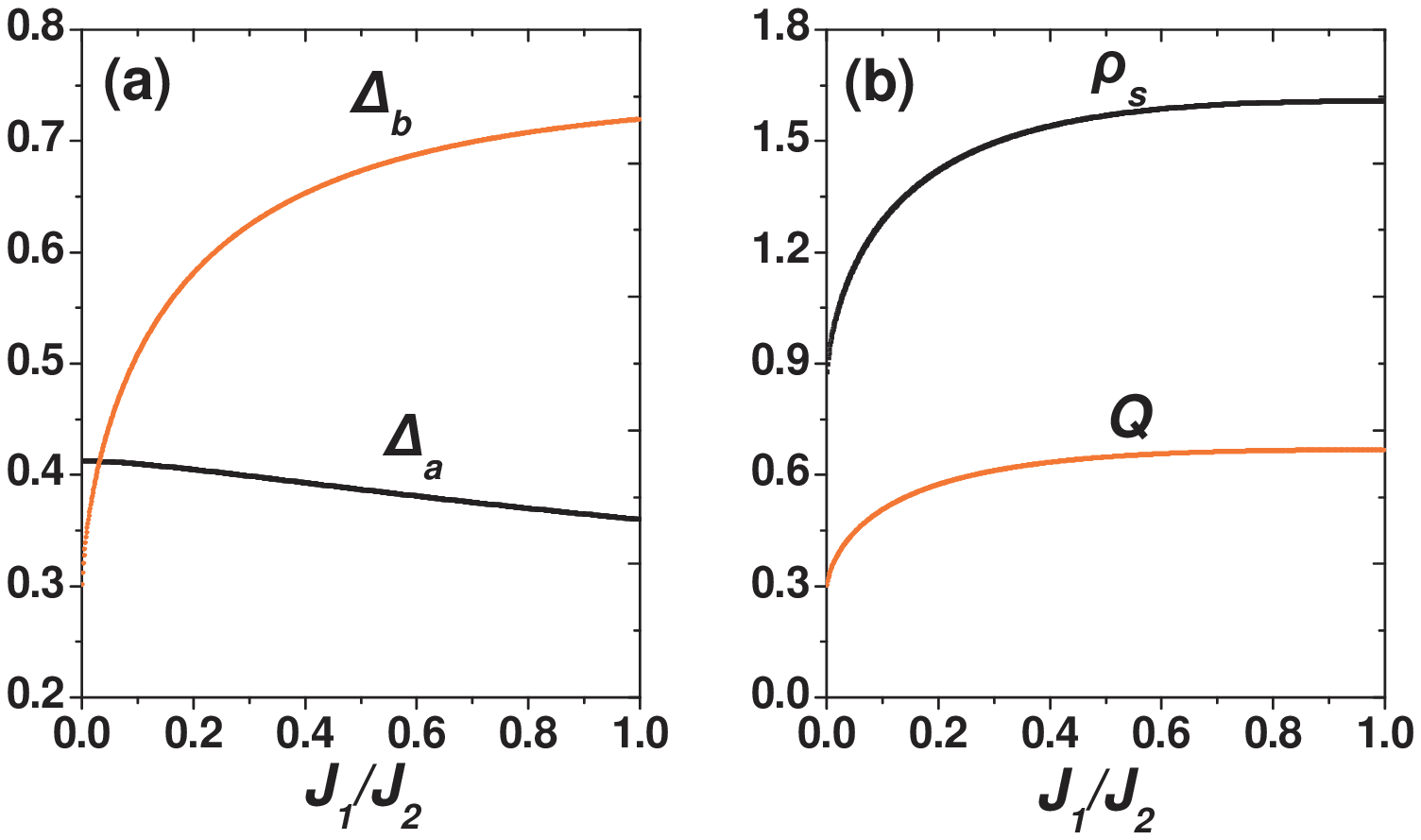}
\end{center}
\caption{Saddle point solution on a 2D square lattice is shown: the valence
bond parameters (a), the superfluid density and quadrupole moment (b) as
functions of $J_1/J_2$. }
\label{fig3}
\end{figure}

It is also more important to examine the spin and quadrupole spatial
correlations in the ordered nematic phase. The calculated static spin
structure factor $S_{D}\left( q_{x},q_{y}\right) $ is shown in Fig.4a, where
a strong broad peak is displayed at $\mathbf{q=(\pi ,\pi )}$, indicating
dominant antiferromagnetic spin correlations, and weaker broad peaks at $%
\mathbf{q=(}0\mathbf{,\pi )}$ and $\mathbf{q=(\pi ,}0\mathbf{)}$,
corresponding to collinear spin-dimer correlations. As a comparison, the
calculated static quadrupole structure factor $S_{Q}\left(
q_{x},q_{y}\right) $ is displayed in Fig.4b, where a $\delta$-like resonance
appears at $\mathbf{q}=(0,0)$, implying strong \textit{ferroquadrupolar}
long-range correlations. These are unique features of an ordered spin
nematic phase, which can be directly probed by polarized inelastic light
scattering experimentally.

\begin{figure}[tbp]
\begin{center}
\includegraphics[width=3.0in]{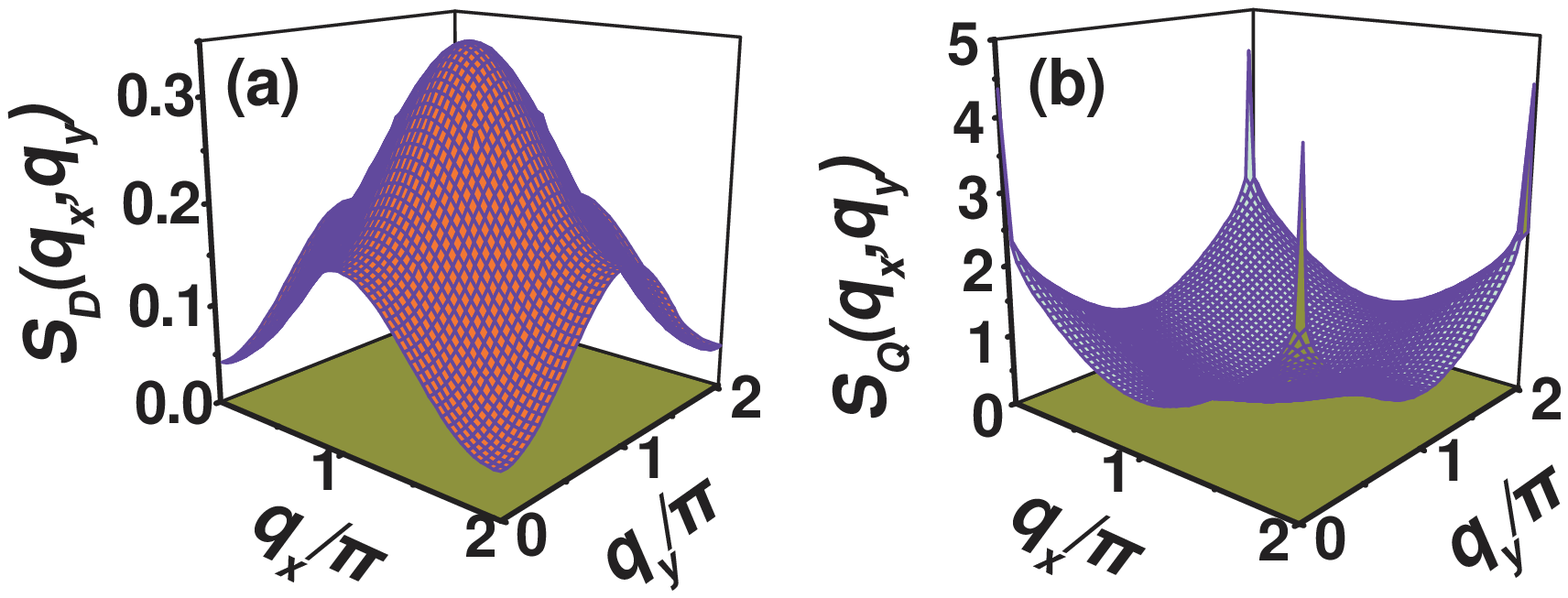}
\end{center}
\caption{Spin structure factor (a) and quadrupole structure factor (b) for
the 2D spin nematic ordered phase with $J_{1}/J_{2}=0.3$.}
\label{fig4}
\end{figure}

To summarize, the SU(3) boson representation is used to develop an efficient
valence bond MF theory for the Mott insulation phase with odd number of
spin-1 bosons in optical lattices. A \textit{first-order} quantum phase
transition from a spin singlet to a nematic phase with gapful excitations is
predicted for 1D, while on a 2D square lattice a spin nematic \textit{%
ordered }phase is shown to always prevail. Both predictions are further
supported by an explicit calculation of the spin and quadrupole correlation
functions.

G.-M. Zhang is grateful to H. -F. L\"{u} for his help in numerical
calculations, and acknowledges the support of NSF-China (No.10125418 and
10474051) and the Special Fund for Major State Basic Research Projects of
China (No.G200067107). L. Yu also acknowledges the support of NSF-China.


\begin{thebibliography}{99}
\bibitem{greiner} M. Greiner \textit{et al}., Nature \textbf{415}, 39 (2002).

\bibitem{jaksch} D. Jaksch \textit{et al}., Phys. Rev. Lett. \textbf{81},
3108 (1998).

\bibitem{duan} L.-M. Duan, E. Demler, and M. D. Lukin, Phys. Rev. Lett.
\textbf{91}, 090402 (2003).

\bibitem{cirac} J. J. Garc\'{\i}a-Ripoll, M. A. Martin-Delgado, and J. I.
Cirac, Phys. Rev. Lett. \textbf{93}, 250405 (2004).

\bibitem{demler-zhou} E. Demler and F. Zhou, Phys. Rev. Lett. \textbf{88},
163001 (2002).

\bibitem{zhou} M. Snoek and F. Zhou, Phys. Rev. B \textbf{69}, 094410 (2004).

\bibitem{demler} A. Imambekov, M. Lukin, and E. Demler, Phys. Rev. A \textbf{%
68}, 063602 (2003); Phys. Rev. Lett. \textbf{93},120405 (2004).

\bibitem{skyip} S. K. Yip, Phys. Rev. Lett. \textbf{90}, 250402 (2003).

\bibitem{ho} T. L. Ho, Phys. Rev. Lett. \textbf{81}, 742 (1998); T. Ohmi and
K. Machda, J. Phys. Soc. Jpn. \textbf{67}, 1822 (1998).

\bibitem{tsuchiya} S. Tsuchiya, S. Kurihara, and T. Kimura, Phys. Rev. A
\textbf{70}, 043628 (2004).

\bibitem{chen-levy} H. H. Chen and P. M. Levy, Phys. Rev. B \textbf{7}, 4267
(1973); N. Papanicolaou, Nucl. Phys. \textbf{B 305}, 367 (1988).

\bibitem{harada} K. Harada and N. Kawashima, Phys. Rev. B \textbf{65},
052403 (2002).

\bibitem{chubukov} A. V. Chubukov, Phys. Rev. B \textbf{43}, 3337 (1991).

\bibitem{solyom} G. F\'{a}th and J. S\'{o}lyom, Phys. Rev. B \textbf{51},
3620 (1995).

\bibitem{batista} C. D. Batista, G. Ortiz, and J. E. Gubernatis, Phys. Rev.
B \textbf{65}, 180402 (2002).

\bibitem{affleck} I. Affleck, Nucl. Phys. B \textbf{265}, 409 (1986).

\bibitem{barber} M. N. Barber and M. T. Batchelor, Phys. Rev. B \textbf{40},
4621 (1989); A. Klumper, Europhys. Lett. \textbf{9}, 815 (1989).

\bibitem{arovas-auerbach} D. P. Arovas and A. Auerbach, Phys. Rev. B.
\textbf{38}, 316 (1988); Phys. Rev. Lett. \textbf{61}, 617 (1988).

\bibitem{gmzhang} G. -M. Zhang \textit{et al.}, in preparation.

\bibitem{kawashima} N. Kawashima, Prog. Theor. Phys. Suppl. \textbf{145},
138 (2002).

\bibitem{lauchli} A. La\"{u}chli, G. Schmid, and S. Trebst, cond-mat/0311082.

\bibitem{ivanov} B. A. Ivanov and A. K. Kolezhuk, Phys. Rev. B \textbf{68},
052401 (2003).
\end{thebibliography}
\end{document}